\documentclass{Interspeech}
\usepackage[utf8]{inputenc} 
\usepackage[T1]{fontenc}    
\usepackage{hyperref}       
\usepackage{url}            
\usepackage{booktabs}       
\usepackage{amsfonts}       
\usepackage{siunitx}
\usepackage{nicefrac}       
\usepackage{microtype}      
\usepackage{graphicx}
\usepackage{doi}
\usepackage{wrapfig}
\usepackage{multicol}
\usepackage{tikz}
\usepackage{pgfplots}
\usepackage{pgf-pie}
\usepackage{graphicx}

\usepackage{multirow}      
\usepackage{booktabs}      
\usepackage{pifont}        
\usepackage{array}         
\usepackage{caption}       
\usepackage{makecell}
\usepackage{marvosym}
\usepackage{hyperref}

\newcommand{\cmark}{\ding{51}}%
\newcommand{\xmark}{\ding{55}}%

\pgfplotsset{compat=1.16}

\definecolor{pie1}{RGB}{ 4 90 141}
\definecolor{pie2}{RGB}{ 241 238 246}
\definecolor{pie3}{RGB}{ 189 201 225}
\definecolor{pie4}{RGB}{ 116 169 207}
\definecolor{pie5}{RGB}{ 43 140 190}

\interspeechcameraready 

\newcommand{\corpusone}{SPGISpeech 1.0}
\newcommand{\corpustwo}{SPGISpeech 2.0}
\newcommand{\spgispeech}{SPGISpeech}
\newcommand{\spglobal}{S \& P Global}
\newcommand{\footnoteone}{https://datasets.kensho.com/datasets/spgispeech2}

\title{\corpustwo{}: Transcribed multi-speaker financial audio for speaker-tagged transcription}

\author[affiliation={1}]{Raymond}{Grossman}
\author[affiliation={2}]{Taejin}{Park} 
\author[affiliation={2}]{Kunal}{Dhawan} 
\author[affiliation={1}]{Andrew}{Titus}
\author[affiliation={1}]{Sophia}{Zhi}
\author[affiliation={1}]{Yulia}{Shchadilova}
\author[affiliation={2}]{Weiqing}{Wang}
\author[affiliation={2}]{Jagadeesh}{Balam}
\author[affiliation={2}]{Boris}{Ginsburg}

\affiliation{}{Kensho Technologies}{USA}
\affiliation{}{NVIDIA Corporation}{USA}
\email{raymond.grossman@kensho.com, taejinp@nvidia.com, kdhawan@nvidia.com}

\usepackage{comment}

\begin{document}
\maketitle
\begin{abstract}
We introduce \corpustwo, a dataset suitable for speaker-tagged transcription in the financial domain. \corpustwo{} improves the diversity of applicable modeling tasks while maintaining the core characteristic of the original \spgispeech{} dataset: audio snippets and their corresponding fully formatted text transcriptions, usable for end-to-end automatic speech recognition (ASR). \corpustwo{} consists of 3,780 additional hours of professionally transcribed earnings calls. Furthermore, the dataset contains call and speaker information for each audio snippet facilitating multi-talker ASR. We validate the utility of \corpustwo{} through improvements in speaker-tagged ASR performance of popular speech recognition models after fine-tuning on \corpustwo{}. Released free for non-commercial use\footnote{\footnoteone{}}, we expect \corpustwo{} to foster advancements in speech recognition technologies and inspire a wide range of research applications.
\end{abstract}

\section{Introduction}
\begin{table*}
\centering
\caption{\textbf{A comparison of speaker recognition corpora.} We compare the size, speaker counts, and style metadata of \corpustwo{} and various peer corpora.}
\begin{tabular}{ r | c c c c } 
\Xhline{2\arrayrulewidth} \toprule
    \textbf{Corpus} &  \textbf{Speaking Style} &  \textbf{Transcription Style} &  \textbf{Speaker Count} &  \textbf{Audio Length(h)} \\ \midrule
    Vox-Celeb1  & Spontaneous, Narrated & None & 1251 & 350  \\
    Vox-Celeb2  & Spontaneous, Narrated & None & 6112 & 2442 \\
    Fisher  & Spontaneous & None & 951 & 980  \\
    Switchboard  & Spontaneous & Orthographic & 543 & 260  \\
    LibriSpeech & Narrated & Non-Orthographic &  28 & 336  \\ 
    \corpustwo{} & Spontaneous, Narrated & Orthographic &  41593 & 3780 \\
\bottomrule \Xhline{2\arrayrulewidth} 
\end{tabular}
\label{table:corpora}
\end{table*}
Automatic speech recognition (ASR) systems have seen increased adoption and deployment in numerous industries during the past decade, primarily due to advances in language modeling and neural network architectures \cite{ puvvada2024moreaccuratespeechrecognition, gulati2020conformer, radford2022robustspeechrecognitionlargescale}. Along with a surge in the adoption of ASR, there has been a growing desire for accurate and robust solutions for speaker related tasks \cite{park2021review, cheng2024integratingaudiovisualsemantic}. The speaker tasks we refer to in this paper are speaker diarization, speaker recognition, and speaker-tagged transcription.  During speaker diarization, a model attempts to allocate portions of spoken audio to generic unknown speakers such as speaker-1 or speaker-2 \cite{1677976}. In contrast, during speaker recognition, a model attempts to allocate portions of spoken audio to specific known speakers based on the characteristics of those speakers \cite{bai2021speakerrecognitionbaseddeep}. Furthermore, speaker-tagged transcription references the integrated tasks of ASR transcription and diarization \cite{park2024sortformerseamlessintegrationspeaker}.

\subsection{Motivation}
Despite significant adoption and demand for robust speaker systems, the datasets available for research on speaker-tagged transcription are minimal compared to their counterparts in ASR transcription. To our knowledge, the most substantial dataset spans two thousand hours of audio \cite{Chung_2018}, while there is an order of magnitude more public data available for ASR transcription \cite{chen2021gigaspeech, oneill2021spgispeech}. Furthermore, many existing speaker datasets suffer from additional complications. These include a lack of orthographically normalized text \cite{Chung_2018}, artificial speaker conditions such as audiobook narration \cite{7178964}, a lack of unknown speakers or sufficient numbers of speakers \cite{cieri2004, godfrey1993}, and limited coverage for characteristics such as gender, ages, or accents \cite{cieri2004, godfrey1993}. Due to these issues, neural speaker diarization models are not as mature and generalizable as those used in speech-to-text transcription \cite{park2021review}. Moreover, most modern neural networks for diarization are not usable or perform worse in an online setting, especially when there are numerous speakers or unknown speakers \cite{park2021review}.
 
In addition, to address increasingly large and complex language and speech tasks, model sizes are likewise increasing \cite{smith2022using}. However, without large and diverse corpora for training, speaker recognition models have been slow to follow this trend. For example, performant ASR systems with integrated speaker solutions are relatively new \cite{park2024sortformerseamlessintegrationspeaker}. To this end, we release \corpustwo{}, a corpus of just under 4,000 hours of audio and the associated speaker-tagged transcriptions. 

\subsection{Features of \corpustwo{}}
Similarly to the original \spgispeech{} \cite{oneill2021spgispeech}, this dataset contains: 
\begin{itemize}
  \item 3,780 hours of audio from corporate earnings calls in English.
  \item Fully-formatted, professional, manual transcription.
  \item Both spontaneous and narrated speech.
  \item Varied teleconference recording conditions.
  \item A diverse selection of L1 and L2 accents, spread globally.
  \item Varied topics relevant to business and finance, including
a broad range of named entities. 
\end{itemize}

Unlike the original \spgispeech{}, \corpustwo{} also offers the following information and structure:
\begin{itemize}
  \item Snippets of 50 to 90 seconds. The original \spgispeech{} contained snippets of up to 15 seconds.
  \item Groupings of snippets from single audio sources containing up to seven speakers and 25 unique speaker segments per snippet. Every snippet in the dataset contains at least two speakers and one speaker change.
  \item 41,593 unique speakers for speaker recognition, many appearing in multiple audio sources.
  \item Per-word speaker information from professional manual transcription.
  \item Alignment-generated per-word timestamps for speaker segments. These alignments were cross-referenced with multiple alignment pipelines.
  \item Algorithmically generated normalized transcriptions. This transcription adjusts how items such as numerical values and disfluencies are represented in the transcript. 
\end{itemize}

It is important to note that the data generation process differed from the original \spgispeech{}, causing more variation in the snippets. Most of these differences arose from two root causes. First, multi-speaker snippets represent a smaller portion of the available pool of data, as earnings call data is biased toward long single-speaker presentations. Second, as we intended to work with significantly longer snippets to enable more robust speaker diarization systems, we needed to relax several constraints around what constituted acceptable snippets from the original \spgispeech{} or too few snippets would be usable.

\section{Prior Work}
\subsection{Previous Speaker Recognition Corpora}	
There are several speaker recognition corpora available, although they vary significantly in terms of size, orthographic style, and speech format. Some commonly used corpora are: Vox-Celeb1 \cite{Nagrani_2017}, Vox-Celeb2 \cite{Chung_2018}, Fisher \cite{cieri2004}, Switchboard \cite{godfrey1993}, and LibriSpeech \cite{7178964}. Information on these corpora can be found in Table~\ref{table:corpora}.

The Vox-Celeb datasets are the most substantial for speaker recognition but do not offer transcription information. The lack of transcriptions limits their usefulness as a general-purpose ASR and speaker recognition dataset. Additionally, data formatting of positive and negative pairs presents a challenge for some speaker-related tasks. In particular, these datasets are less applicable to speaker-tagged transcription and speaker diarization, especially where the speaker counts and durations are unknown \cite{Chung_2018, Nagrani_2017}.

In contrast, the Fisher dataset offers transcriptions, but consists solely of telephone calls between a limited number of speakers. This setup reduces the variance in audio conditions and speaker characteristics \cite{cieri2004}. Similarly, the Switchboard datasets offer a limited number of speakers over a set of audio an order of magnitude smaller than its Vox-Celeb counterparts \cite{godfrey1993}.

Finally, the LibriSpeech corpus, a standard benchmark for ASR and speaker recognition models, consists solely of narrated audiobooks in the public domain. Narrated speech lacks the acoustic characteristics found in spontaneous speech, reducing the generalizability of models trained on the dataset \cite{7178964}.

It is important to note two additional issues. First, many of these datasets are not regularly updated to capture dialectical shifts over time. Additionally, many speaker recognition corpora are locked behind paywalls and require an investment before use in research. 

\corpustwo{} is the first corpus free for non-commercial use offering a dataset containing many speakers over a wide range of demographics and orthographic text transcription. Although it is more substantial in scope and speaker offerings than the previous datasets, there are still limitations within the dataset. The topics are primarily finance-related, leading to a domain-specific corpus. Additionally, the speaker demographic skews towards individuals likely to be present on financial calls, rather than a balanced, representative sample of the population. Finally, there is no human-annotated time alignment for words and their location in the audio snippets.

\begin{table*}
\centering
\caption{\textbf{A comparison of the distribution of entities and interesting textual features between the original \spgispeech{} dataset and \corpustwo{}.} Percentages represent the ratio of snippets with the feature to the total number of snippets. As the \corpustwo{} snippets are around 5x longer, we expect more occurences per snippet.}
\begin{tabular}{ r | c c c c c } 
\Xhline{2\arrayrulewidth} \toprule
    \textbf{Corpus} & \textbf{Acronyms(\%)} & \textbf{Pauses(\%)} & \textbf{Organizations(\%)} & \textbf{Persons(\%)} & \textbf{Locations(\%)} \\ \midrule
    \spgispeech{} & 15 & 10 & 25 & 8 & 8 \\
    \corpustwo{} & 23 & 66 & 52 & 45 & 23 \\ \bottomrule \Xhline{2\arrayrulewidth}
\end{tabular}
\label{table:textual_features}
\end{table*}

\begin{figure}
\centering
\begin{tikzpicture}[scale=.93]
\begin{semilogyaxis}[ybar interval,
    ymin=1, ymax=210000, domain=1:180000,
    xtick style={draw=none}, ytick style= {draw=none}, 
    area style,
    ylabel= \# of Snippets,
    xlabel= Speaker Count,
    height=120pt,
    width = 230pt,
    grid=none]

\addplot [fill=pie1, ybar interval,  mark=no] plot coordinates{(2, 142736) (3, 30289) (4, 2648) (5, 153) (6, 9) (7, 1)};

\end{semilogyaxis}
\end{tikzpicture}
\caption{\textbf{Distribution of the number of unique speakers per snippet}. There is also a single snippet with more than six speakers.}
\label{figure:unique_speakers}
\end{figure}
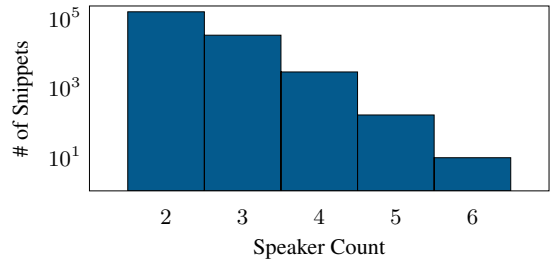

\subsection{Major Differences to \corpusone{}}	
 We relaxed two constraints used in creating \corpusone{} during the data generation process for \corpustwo{}. Those constraints were the removal of snippets with names and the removal of snippets that follow the \spglobal{} styling guidelines instead of common ASR transcription norms. In the creation of the \corpusone{} dataset, audio snippets containing any occurrences of names, numerals, or significant style guide issues were removed from the pool of candidate snippets. However, this was not possible for \corpustwo{}. Unlike in \corpusone{}, in \corpustwo{} snippets were mainly sourced from the messier Q \& A portions of calls where names and style guide issues are more prevalent. Furthermore, the target snippet length quintupled from \corpusone{} to \corpustwo{}, further exacerbating this challenge.

 The inclusion of snippets that violate these constraints adds additional complexity to the dataset because the transcription is not always literal. The style guide \spglobal{} uses for professional transcription usually results in the transcription of the intended content of the speech, rather than a verbatim transcription of the speaker. For example, if a speaker stutters before correctly saying a word, only the final correct word will be transcribed under the \spglobal{} style guide. Similarly, certain smalltalk phrases irrelevant to the business purpose of the call are deleted, and filler words and disfluencies likewise removed as they do not add to the business context. Several of the commonly edited phrases are "Hey", "Thanks/Thank you", and variations on "Uh/Um/Hmm/Like".

 To address these issues, we have provided both the human-annotated transcription and an algorithmically adjusted version that attempts to modify areas of the calls where numeric and style guide issues occur. The adjusted version aims to provide a more literal transcription of the audio. This adjustment process consisted of two parts. First, we ran an ASR model \cite{puvvada2024moreaccuratespeechrecognition} trained on literal transcriptions across the entire dataset and aligned the resulting text with the gold label annotations. Then, numeric values were cross-referenced with model transcriptions to attempt to identify and substitute the correct denormalization of numeric values. Additionally, disfluencies, removed small talk, and removed stutters/corrections appearing in the model transcription were identified and reintroduced where possible for a more literal transcription of the audio.

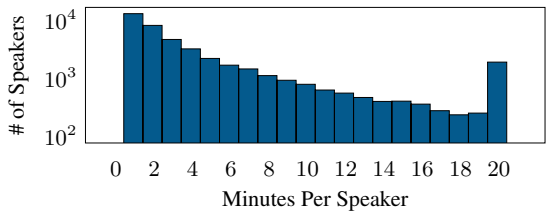
\begin{figure}
\centering
\begin{tikzpicture}[scale=.93]
\begin{semilogyaxis}[ybar interval,
    ymin=0, ymax=16000, 
    minor xtick={1, 3, 5, 7, 9, 11, 13, 15, 17, 19},
    xtick={0,2,4,6,8,10,12,14,16,18,20,22},
    xtick pos=bottom, ytick style= {draw=none}, 
    area style, ylabel= \# of Speakers,
    xlabel= Minutes Per Speaker,
    x tick label style={black, xshift=1.6mm},
    minor tick style={draw=none},
    major tick style = {draw=none},
    grid=none,
    height=100pt,
    width = 230pt
    ]
\addplot [fill=pie1, ybar interval,mark=no] plot coordinates{
(2, 12469)
(3, 7832)
(4, 4481)
(5, 3082)
(6, 2110)
(7, 1615)
(8, 1390)
(9, 1067)
(10, 887)
(11, 756)
(12, 602)
(13, 533)
(14, 448)
(15, 383)
(16, 387)
(17, 344)
(18, 265)
(19, 225)
(20, 241)
(21, 1823)
(22, 120)};

\end{semilogyaxis}
\end{tikzpicture}
\caption{\textbf{Distribution of number of minutes of speech for each unique speaker}. Rounded up. Due to the long tail of the distribution, the final bucket contains all speakers who talk for 20 minutes or more.}
\label{figure:speaker_minutes}
\end{figure}

\section{Corpus Generation}
 
The \corpustwo{} corpus resembles the original \spgispeech{} corpus in many ways. Both consist of snippets derived from professionally transcribed earnings calls. Additionally, the transcriptions follow the same style guide as the original dataset, meaning orthographic conventions are identical between the two datasets. These conventions include capitalization, punctuation, text normalization, and transcription of disfluencies.

Additionally, the file types for the ASR portion of the datasets are identical. All snippets are 16-kHz single-channel audio files. The alignment and slicing process, using the Gentle alignment package \cite{Gentle} and py-webrtc \cite{pyWebRTC} for voice activity detection (VAD), also partly matches the original \spgispeech{} \cite{oneill2021spgispeech}. However, these alignments were cross-referenced with the NeMo forced alignment pipeline \cite{Harper_NeMo_a_toolkit} to ensure accurate word-level alignments, particularly at the start and end of calls. 

Despite these similarities, there are significant differences between the two corpora. \corpustwo{} contains significantly longer audio snippets of 50-90 seconds in length to support the development of diarization systems for longer calls. Additionally, each audio file also has an associated speaker alignment file in the SegLST format as seen in the CHiME challenges \cite{cornell2024chime8dasrchallengegeneralizable}, as well as algorithmically generated word-level alignments in JSON format. Finally, speaker information has been included in both the transcription and in the metadata. In the transcripts, every occurrence of a speaker change is represented by the pipe character "|".  The speakers for each segment and word can be found in the SegLST files and in the JSON word-level alignments, respectively. Moreover, speakers are assigned a random anonymous positive integer ID that is consistent across calls, or "-1" for unknown speakers.

\subsection{Sample Selection and Dataset Characteristics}


Several additional criteria determined the inclusion or exclusion of a snippet from the data set, the first three of which were adapted from the original \spgispeech{}. 

\begin{enumerate}
	\item Few non-standard characters or currency symbols.
	\item Snippets in the dataset could consist of no more than 35\% of the audio of a given call.
	\item Snippets were required to be well-aligned for a minimum of five words at the start and end of each snippet under both the Gentle alignment framework and the NeMo forced alignment framework. 
	\item Snippets with known speakers were prioritized over snippets with unknown speakers.
	\item At least 50\% of the speakers in dev and test appear in train.
        \item Both a mean and a median of at least four snippets per speaker.
\end{enumerate}

\begin{table*}[h]
    \centering
    \caption{\textbf{Performance comparison of different models and configurations.} For definitions of PnC, WER, and cpWER, see section 5.2.}
    \label{table:model_results}
    \setlength{\tabcolsep}{4.0pt} 
    \renewcommand{\arraystretch}{1.1}
    \begin{tabular}{cccc|cc|cc}
    \Xhline{2\arrayrulewidth} \toprule
        & \textbf{ASR} & \textbf{Speaker} & \textbf{Fine-tuned On} & \multicolumn{2}{c|}{\textbf{With PnC}} & \multicolumn{2}{c}{\textbf{Without-PnC}} \\
        \textbf{S.No.} & \textbf{Model} &\textbf{Supervision} & \textbf{\corpustwo}  & \textbf{cpWER\%} & \textbf{WER\%} & \textbf{cpWER\%} & \textbf{WER\%} \\
        \midrule
        baseline-1 & Canary-170M & \xmark & \xmark & - &  10.81 & - & 6.48 \\
baseline-2 & Canary-1B & \xmark & \xmark & - &  10.32 & - & 5.87 \\
        baseline-3 & Whisper-turbo & \xmark & \xmark & - &  24.77 & - & 19.10 \\
        baseline-4 & Whisper-large-v3 & \xmark & \xmark & - &  18.31 & - & 12.46 \\
        \midrule
        1 & Canary-170M & \xmark & \cmark &  20.53 & 7.96 & 18.08 & 5.33 \\
        2 & Canary-170M + Sortformer-123M & \cmark & \cmark &  \textbf{15.88} & \textbf{7.25} & \textbf{13.39} & \textbf{4.62} \\
        \bottomrule
        \Xhline{2\arrayrulewidth}
    \end{tabular}
    \vspace{-2ex}
\end{table*}

\subsection{Dataset Split}

The dataset is divided into a train, dev, and test set. The train set consists of 154,971 snippets spanning 48,000 earnings calls, the dev set consists of 6,447 snippets spanning 2,000 earnings calls, and the test set consists of 7,177 snippets spanning 2,131 earnings calls. For any given call, the snippets will appear in only one of the splits. Moreover, the dev and test sets consist of the most recent calls chronologically. All calls in the training set occur before any calls in the dev set, and all calls in the dev set occur before any calls in the test set. This ensures no leakage of information chronologically between train, dev, and test. 

\begin{figure}
\centering
\begin{tikzpicture}
\def\printonlylargeenough#1#2{\unless\ifdim#2pt<#1pt\relax
#2\printnumbertrue
\else
\printnumberfalse
\fi}
\newif\ifprintnumber
\pie[text=legend,radius=1.6,rotate=180,color={pie1, pie5, pie4, pie3, pie2}, before number=\printonlylargeenough{4}, after number=\ifprintnumber\%\fi]{49/North America, 24/\text{Asia/Pacific}, 20/Europe, 5/\text{Latin America/Caribbean}, 2/\text{Africa/Middle East}}
\end{tikzpicture}
\caption{\textbf{Distribution of speakers by region}. Estimated based on corporate domicile location for 2,000 random speakers.}
\label{figure:speaker_locations}
\end{figure}
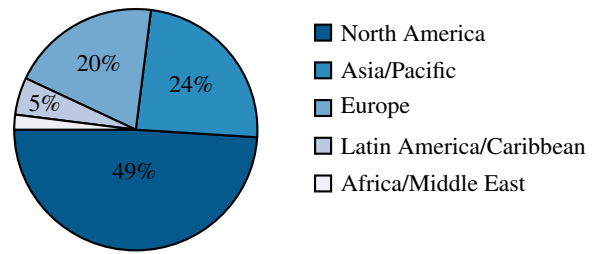

\subsection{Overlapping and Low Quality Speech}
Overlapping speech remains one of the main challenges for ASR and speaker recognition systems \cite{park2021review}. Overlapping speech in our dataset is less prevalent than in other speech domains due to the professional nature of earnings calls. According to the style guide used by human annotators, unless the overlapped speech is unintelligible, the transcription will be split by speaker. The first speaker will be transcribed entirely before the second speaker is transcribed. Additionally, we attempted to remove the most egregious overlapped and low-quality snippets from the dataset. Earnings calls on which our preexisting ASR, diarization, and speaker recognition models performed in the bottom decile were excluded from the dataset.

\section{Corpus Characteristics}
The dataset consists of 3,780 hours of audio snippets and more than 150,000 unique individual snippets between 50 and 90 seconds long, with an average snippet duration of 77.3 seconds. There are more than 40,000 unique speakers, with an average of 8.1 snippets per speaker in the dataset (ignoring unknown speakers). To our knowledge, this is the largest speaker recognition dataset available both by number of hours and number of speakers. 

The dataset is similar to the original \spgispeech{} demographically and in the prevalence of entities and acronyms, as can be seen in Table~\ref{table:textual_features}. To maintain a fair comparison, the prevalence of people, organizations, and locations was again estimated with Flair NER \cite{akbik-etal-2018-contextual} on a subset of 2,500 snippets. Furthermore, both the original and 2.0 datasets contain information from all eleven top-level Global Industry Classification Standard (GICS) sectors \cite{gics} and, as such, contain a comprehensive subset of modern English business discourse. Moreover, the dataset contains a global distribution of speakers, as can be seen in Figure~\ref{figure:speaker_locations}. 

The distribution of the number of unique speakers per snippet and the distribution of the number of minutes per speaker can be found in Figure~\ref{figure:unique_speakers} and Figure~\ref{figure:speaker_minutes}, respectively. In general, snippets have relatively few speakers due to the professional nature of earnings calls, but most speakers appear in numerous snippets, with a mean of just over 5 minutes of audio per speaker.

\section{Benchmarks}
We train a Canary-based ASR model and a Sortformer-based model for integrated speaker diarization and ASR transcription to demonstrate the serviceability of \corpustwo{} on both transcription and speaker tasks.

\subsection{Model Description and Methods}
Canary and Sortformer are both encoder-decoder based architectures. The original Sortformer uses Canary as the base ASR model with modifications, as does the model which we trained on \corpustwo{}. Canary itself uses a FastConformer encoder \cite{rekesh2023fastconformerlinearlyscalable} and a Transformer decoder \cite{hrinchuk2020correction}. Sortformer adds a parallel encoder to the Canary decoder for speaker diarization, and introduces sort loss to deal with the resolution of speaker permutations from permutation-invariant loss during training \cite{park2024sortformerseamlessintegrationspeaker}. This architecture enables end-to-end training for both diarization and ASR simultaneously, and demonstrates strong performance on existing datasets \cite{park2024sortformerseamlessintegrationspeaker} and on \corpustwo{}.

\subsection{Results and Discussion}

Table~\ref{table:model_results} presents the performance comparison of different ASR models with various configurations. The Canary-170M model, when fine-tuned with speaker supervision and on \corpustwo{}, achieves a significant gain from the baseline ASR models. To show the accuracy on speaker-ID (speaker tagging), we use Concatenated Minimum Permutation Word Error Rate (cpWER)~\cite{watanabe2020chime}, which can reflect speaker-tagging accuracy in the form of Word Error Rate (WER). We define WER in a standard fashion as substitutions, deletions, and insertions divided by the number of reference words. cpWER is then calculated by finding the optimal permutation of predicted speaker segments to reference transcripts across all possible permutations, reporting the minimum WER achieved from the optimal permutation.

The results indicate that models without fine-tuning or speaker supervision generally perform worse. The baseline models (Canary-170M, Canary-1B, Whisper-turbo, and Whisper-large-v3) exhibit a higher cpWER and WER in both the With-PnC (Punctuation and Capitalization) and Without-PnC settings. Among them, Whisper-turbo shows the highest WER in the Without-PnC case, while Canary-170M achieves the lowest WER when fine-tuned with speaker supervision \cite{park2024sortformerseamlessintegrationspeaker} on \corpustwo{}.

Moreover, a significant performance boost is observed when incorporating speaker supervision during training on \corpustwo{}. The Canary-170M model with these enhancements achieves 15.88\% cpWER and 7.25\% WER under With-PnC, significantly outperforming the baseline models. These results highlight the performance gain obtained from our proposed dataset and the benefits from speaker-aware training. 


\section{Conclusion}
In this paper, we introduce \corpustwo{}, a 3,780-hour data set for speaker-tagged transcription. This dataset is suitable for both fully formatted ASR transcription and speaker tasks including speaker-tagged transcription, speaker diarization, and speaker identification. We demonstrate the feasibility of these tasks by training an end-to-end model for speaker-tagged transcription on the dataset. Finally, as a contribution to the research community, the dataset is released for non-commercial use.

\bibliographystyle{IEEEtran}
\bibliography{refs}

\end{document}